\documentclass[conference]{IEEEtran}

\usepackage{cite}
\usepackage{amsmath,amssymb,amsfonts}

\usepackage{xcolor}
\usepackage{tikz}
\usetikzlibrary{arrows,fit,positioning}
\usetikzlibrary {shapes.geometric}
\usepackage{pgfplots}
\usepackage{stmaryrd}
\usepackage{listings}
\usepackage{mathtools}
\usepackage{subcaption}
\usepackage{soul}
\pgfplotsset{width=11cm,compat=1.9}
\def\BibTeX{{\rm B\kern-.05em{\sc i\kern-.025em b}\kern-.08em
    T\kern-.1667em\lower.7ex\hbox{E}\kern-.125emX}}

\begin{document}

\title{Automating Constraint-Aware Datapath Optimization using E-Graphs}

\author{\author{\IEEEauthorblockN{Samuel Coward}
\IEEEauthorblockA{Numerical Hardware Group\\
Intel Corporation\\
Email: samuel.coward@intel.com}
\and
\IEEEauthorblockN{George A.~Constantinides}
\IEEEauthorblockA{Electrical and Electronic Engineering\\
Imperial College London\\
Email: g.constantinides@imperial.ac.uk}
\and
\IEEEauthorblockN{Theo Drane}
\IEEEauthorblockA{Numerical Hardware Group\\
Intel Corporation\\
Email: theo.drane@intel.com}}
}

%\author{\author{\IEEEauthorblockN{Anon.}
%\IEEEauthorblockA{Anon.\\
%Anon.\\
%Anon.}
%}}
\newcommand\ass{
    \texttt{ASSUME}
}

\newcommand\abstraction[1]{
    \mathcal{A}\llbracket {#1} \rrbracket
}
\newcommand\lsub[1]{
    \prescript{}{#1}
}

\maketitle
\begin{abstract}
% OLD ABSTRACT
%Numerical hardware design calls for aggressive levels of optimization, forcing designers to understand deep properties of their design. A powerful technique is case splitting, introducing conditional branches in the design. The designer is able to exploit the fact that a branch is only taken under some set of conditions, creating optimization opportunities that are valid only on a sub-domain of the input space. We have developed an RTL optimization tool that automatically learns the consequences of conditional branches and exploits that knowledge to enable deep optimization for substantial performance gains. The tool deploys a custom built program analysis based on the theory of abstract interpretation, which when combined with a data-structure known as an e-graphs can simplify complex reasoning about program properties. To demonstrate the potential, we describe an e-graph based tool to optimize a naive floating-point circuit design. Our tools is able to fully-automatically discover known floating-point architectures from the computer arithmetic literature. The tool is applied to a number of numerical RTL designs and out-performs baseline EDA tools, which do not possess these deeper analysis capabilities. \gc{Some kind of quantitative summary here?}

Numerical hardware design requires aggressive optimization, where designers exploit branch constraints, creating optimization opportunities that are valid only on a sub-domain of input space. We developed an RTL optimization tool that automatically learns the consequences of conditional branches and exploits that knowledge to enable deep optimization. The tool deploys custom built program analysis based on abstract interpretation theory, which when combined with a data-structure known as an e-graph simplifies complex reasoning about program properties. Our tool fully-automatically discovers known floating-point architectures from the computer arithmetic literature and out-performs baseline EDA tools, generating up to 33\% faster and 41\% smaller circuits.
\end{abstract}

% \keywords{datapath design, elementary function, polynomial interpolation}

\maketitle

%%%%%%%%%%%%%%%%%%%%%%%%%%%%%%%%%%%%%%%%%%%%%%%%%%%%%%%%%%%%%%%%%%%%%%%%
% INTRODUCTION
%%%%%%%%%%%%%%%%%%%%%%%%%%%%%%%%%%%%%%%%%%%%%%%%%%%%%%%%%%%%%%%%%%%%%%%%
\section{Introduction}
Industrial Register Transfer Level (RTL) design requires engineers to exploit all possible optimization opportunities in increasingly complex designs. In particular, floating-point hardware is the subject of intense scrutiny given its wide ranging applications. These modules are almost always designed by hand~\cite{Beaumont-Smith1999ReducedArchitectures, Sohn2012ImprovedUnit,Farmwald1981OnUnits}. State-of-the-art electronic design automation (EDA) tools are currently unable to match human designs.

Combining program analysis and e(quivalence) graph rewriting techniques we exceed the optimization capabilities of existing EDA tools for hardware design, with the ability to automatically exploit optimization opportunities generated by conditional branches in designs.
We will first discuss key topics and prior work. In Section \ref{sect:theory} we will introduce the theory underpinning sub-domain equivalences and how conditional branches are captured in e-graphs to compute tight approximations to intermediate program values. In Section \ref{sect:implementation} we will discuss how constraint-aware optimization can be implemented in an RTL optimization tool, built on top of the \texttt{egg} e-graph library. 
We will then consider a floating-point subtractor design case study in Section \ref{sect:case_study}, with further results in Section \ref{sect:results}.
The paper contains the following novel contributions:
\begin{itemize}
    \item expressibility of sub-domain equivalences in an e-graph to enable constraint-aware datapath optimization,
    \item a method and tool to automate the production of run-time branching / muxing conditions, leading to optimized hardware via constraint-aware optimization,
    \item a case study demonstrating the automated production of a highly-optimized floating-point subtractor, 
    \item evaluation on benchmarks showing the generality of the method.
\end{itemize}
%%%%%%%%%%%%%%%%%%%%%%%%%%%%%%%%%%%%%%%%%%%%%%%%%%%%
% BACKGROUND
%%%%%%%%%%%%%%%%%%%%%%%%%%%%%%%%%%%%%%%%%%%%%%%%%%%%
\section{Background}
\begin{figure}
    \centering
    \begin{subfigure}{.45\columnwidth}
    \includegraphics[scale=0.44]{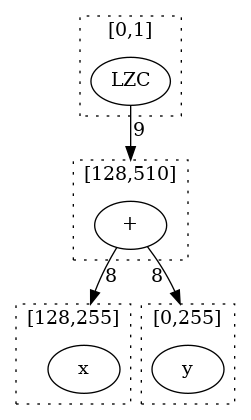}
    \caption{Initial e-graph represents $\textrm{LZC}(x+y)$.}
    \label{fig:initial_lzc}
    \end{subfigure}\quad%
\begin{subfigure}{.45\columnwidth}
  \centering
  \includegraphics[scale=0.44]{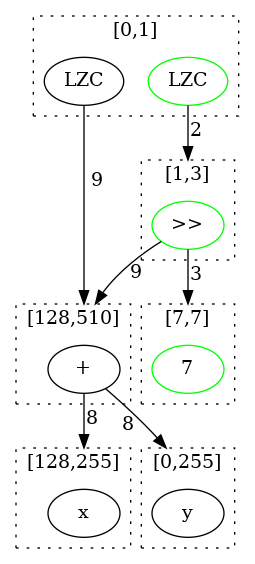}
    \caption{$\textrm{LZC}(a)\rightarrow \textrm{LZC}(a \gg 7)$. Green nodes are newly added.}
    \label{fig:final_lzc}
\end{subfigure}
\caption{An e-graph before and after rewriting. The LZC node denotes a leading-zero counter. Edges are labelled with bitwidths. The rewrite is valid due to the input constraint, $x\geq 128$, which implies that LZC($x+y$)$\leq 1$, namely $x+y$ has at most one leading zero.}
\label{fig:lzc_diagram}

\end{figure}

E-graphs are a data structure that represents equivalence classes (e-classes) of expressions compactly~\cite{Nelson1980TechniquesVerification,Willsey2021Egg:Saturation}. Nodes in the e-graph represent functions or arguments which are grouped into e-classes. Directed edges connect a node to its child e-classes, representing the function's inputs. An e-graph is grown via the application of rewrites (e.g. $\texttt{x+0} \rightarrow \texttt{x}$), which define equivalences over expressions. Constructive rewrite application means that the left-hand side remains in the data structure after application, avoiding the phase ordering problem~\cite{Tate2009EqualityOptimization}. The e-graph grows monotonically, representing an increasing design space of equivalent implementations. An example e-graph before and after rewriting is presented in Figure \ref{fig:lzc_diagram}, which also shows how interval enclosures can be attached to each e-class from which we can learn useful properties \cite{Willsey2021Egg:Saturation, Coward2022CombiningInterpretation}. It is possible to learn such properties via abstract interpretation as we will show in Section \ref{sect:theory}.

Compilers use program analysis to enable optimizations such as dead code elimination~\cite{Cooper2011EngineeringEdition, Lattner2004LLVM:Transformation}. Abstract interpretation is a theory used to over-approximate program properties ~\cite{Cousot1977AbstractFixpoints}. A weak but cheap to compute interpretation is interval arithmetic, which approximates variables by their input ranges and operators by their natural interval extension. Relational domains such as the polyhedral domain~\cite{Chen2008ADomain} capture correlation and are more powerful but are more complex to compute.

On the datapath optimization problem, syntactic rewriting techniques have been explored~\cite{dataflow2008verma}, with one contribution using e-graphs~\cite{Coward2022AutomaticE-Graphs}. Unfortunately, syntactic rewrites alone cannot express the deep transformations required for floating-point hardware design. We will show that such transformations depend upon knowledge of intermediate values and domain restrictions under which the branches are executed.  
%%%%%%%%%%%%%%%%%%%%%%%%%%%%%%%%%%%%%%%%%%%%%%%%%%%%
% THEORY
%%%%%%%%%%%%%%%%%%%%%%%%%%%%%%%%%%%%%%%%%%%%%%%%%%%%
\section{Theory} \label{sect:theory}
%%%%%%%%%%%%%%%%%%%%%%%%%%%%%%%%%%%%%%%%%%%%%%%%%%%%
% SUB-DOMAIN EQUIVALENCES
%%%%%%%%%%%%%%%%%%%%%%%%%%%%%%%%%%%%%%%%%%%%%%%%%%%%
\subsection{Sub-Domain Equivalences} \label{subsect:sub_domain equivs}
E-graphs represent expressions, drawn from a set \textit{Expr} with variables evaluated over a domain $\mathcal{D}$. We consider a concrete semantics of expressions $\llbracket \cdot \rrbracket_\cdot \in \textit{Expr} \to \mathcal{D}^n \to \mathcal{D}$, evaluating an expression as a function, e.g. $\llbracket \mathtt{x + 1} \rrbracket$ is a function mapping values of the variable $\mathtt{x}$ to values of the expression $\mathtt{x+1}$. This allows us to say that two expressions, $e_a$ and $e_b$, are congruent, $e_a \cong e_b$, iff $\llbracket e_a \rrbracket = \llbracket e_b \rrbracket$.

This notion of congruence is strict and enforces equivalence across the entire domain $\mathcal{D}^n$. 
Under a weaker congruence relation, e.g. equivalence on a sub-domain, many additional congruences may hold. Expressions, $e_a$ and $e_b$, can be said to be {\em congruent under} $c\in \textit{Expr}$ iff 
\[\forall v \in \mathcal{D}^n,\; \llbracket c \rrbracket(v) \Rightarrow \llbracket e_a \rrbracket(v) = \llbracket e_b \rrbracket(v)\]
{\em i.e.} in every model where $\llbracket c\rrbracket$ holds, $\llbracket e_a \rrbracket = \llbracket e_b \rrbracket$. We denote this $e_a \cong_c e_b$.

Input constraints or conditional branches can constrain the domain of operands for a given operator, potentially exposing additional optimization opportunities. The leading zero counter (LZC) circuit described in Figure \ref{fig:lzc_diagram}, demonstrates how an input constraint can be exploited to build an equivalent but smaller and faster circuit.

Conditional branches initiated via an if statement in software or a mux in hardware will be of particular interest in this work. As an example, the following C expressions are equivalent, even though $\text{fabs}(x) \neq x$ in general.
\begin{align*}
 \left(\texttt{x > 0 ? fabs(x) : 0}\right)\; \cong \;\left(\texttt{x > 0 ? x : 0}\right)
\end{align*}
Capturing these possibilities within datapath design optimization is essential. Sub-domain equivalence relations allow developers to optimize each branch of their code under different conditions, often resulting in better performance. This is exactly the motivation behind introducing case splits into any design. Examples can be seen in Sections \ref{sect:case_study} and \ref{sect:results}.

%%%%%%%%%%%%%%%%%%%%%%%%%%%%%%%%%%%%%%%%%%%%%%%%%%%%
% RELATIONAL DOMAINS COMBINED WITH E-GRAPHS
%%%%%%%%%%%%%%%%%%%%%%%%%%%%%%%%%%%%%%%%%%%%%%%%%%%%
\subsection{Sub-Domains in E-Graphs} 
\label{subsect:relation_domains_egraphs}
The preceding examples show the necessity to capture the set of possible variable values under which one cares about the correct evaluation of an expression, and to do so requires automated reasoning about the set of values an expression can take. Abstract interpretation is a well-known approach to this problem in the field of program analysis~\cite{Cousot2021PrinciplesInterpretation}. In our work, we abstract the set of `care' values of an expression as a finite union of integer intervals, as this is sufficient to be able to reason about the industrial datapath designs we encounter. To be more precise, with each expression we associate an element of the set $\mathcal{A}$:
\[\mathcal{A} = \left\{ \bigcup_{i=1}^n [a_i, b_i] \, | \, a_i\leq b_i, \, a_i, b_i \in \mathbb{Z}, \, n \in \mathbb{N} \right\}. \]
For a given $e\in \textit{Expr}$, we compute $\mathcal{A}\llbracket e \rrbracket \in \mathcal{A}$, using interval arithmetic extended to unions of intervals incurring additional computational complexity. Following the technique in~\cite{Coward2022CombiningInterpretation}, each e-class of expressions is associated with an element of $\mathcal{A}$, which represents a conservative approximation of all evaluations of that e-class, as shown in Figure \ref{fig:lzc_diagram}. Such an approach is shown to produce a more accurate program analysis~\cite{Coward2022CombiningInterpretation}. 
%All expressions in an e-class are equivalent therefore evaluate to the same result given identical inputs. Taking the intersection across all node abstractions in an e-class generates tighter abstract approximations, resulting in a more accurate program analysis~\cite{Coward2022CombiningInterpretation}. Such program analysis enables deeper optimizations but does not capture constrained branch execution. 
%\gc{I think the preceding para will be hard for DAC readers to understand what's going on. Can we just wave towards our arXiv paper on combining egraphs with AI and say something like `following the technique in [?], each equivalence class of expressions is associated with an element of $\mathcal{A}$' and then get rid of much of the preceding stuff?}

% \begin{align*}
% \langle \textit{Constr}\rangle ::&= \langle \textit{Expr}\rangle \hspace{0.36em} % <\hspace{0.36em}\langle \mathrm{Const}\rangle\\
% &|\quad \langle \textit{Expr}\rangle \hspace{0.36em} >\hspace{0.36em}    \langle % \mathrm{Const}\rangle \\
% &|\quad \langle \textit{Expr}\rangle ==   \langle \mathrm{Const}\rangle \\
% &|\quad \langle \textit{Expr}\rangle \hspace{0.36em} \neq\hspace{0.36em} \langle % \mathrm{Const}\rangle
% \end{align*}
We introduce an additional operator, $\texttt{ASSUME}$, that will be used to encode sub-domain equivalences. $\texttt{ASSUME}$ takes two operands, an e-class containing equivalent \textit{Expr} to evaluate and a set of e-classes containing \textit{Expr}, encoding conditions that may be assumed to be true when evaluating the first argument. We achieve this effect by appending an additional \textit{special} element to $\mathbb{Z}$, forming a new domain $\mathbb{Z}' = \mathbb{Z} \cup \{*\}$. Naturally extending $\llbracket \cdot \rrbracket$ notation to e-classes, the semantics under a single constraint are:
\begin{equation}
    \llbracket \texttt{ASSUME}(x,c) \rrbracket =        
    \begin{cases}
       \llbracket x \rrbracket &\quad\text{if } \llbracket c \rrbracket\\
       \hspace{0.4em}* &\quad \text{else.} \\
     \end{cases}
\end{equation}
Under multiple constraints, a * is returned if any of the conditions do not hold. The semantics of all other functions except the ternary operator $\cdot ? \cdot : \cdot$ are extended to this new domain by also returning * iff at least one of their operands is *. The ternary operator requires special treatment as it returns a * only if the branching condition itself is a * or one of the {\em reachable} branches returns a *. In this way, we can think of * as precisely capturing the code `failing an assertion'.
The key benefit of this construction is that we may now define $\cong_c$ in terms of congruence over the whole domain, that is
\begin{equation}
    x \cong_c y \Leftrightarrow \texttt{ASSUME}(x,c) \cong \texttt{ASSUME}(y,c).
\end{equation}
Thus we may reason automatically about sub-domain congruences using the e-graph machinery which applies for whole-domain congruences.

The consequences of the constraints are realized in the abstraction of $\texttt{ASSUME}$, if the assumed conditions constrain the expression under evaluation. The abstraction of $\texttt{ASSUME}$ with a single constraint is:
\begin{equation}
    \mathcal{A}\llbracket\texttt{ASSUME}(x, c) \rrbracket = \mathcal{A}\llbracket x \rrbracket \cap I,\textrm{ where}
\end{equation}
\begin{equation} \label{eqn:assume_abstraction_case}
    I =  \\
        \begin{cases}
       (-\infty, c') &\quad\text{if } [x<c']\hspace{0.8em}\in c\\
       (c', \infty) &\quad\text{if } [x>c']\hspace{0.8em}\in c\\
       [c', c'] &\quad\text{if }      [x==c']\in c\\
       (-\infty, c') \cup (c', \infty) &\quad\text{if } [x\neq c']\hspace{0.8em} \in c\\
       (-\infty, \infty) &\quad \text{else.} \\
     \end{cases}
\end{equation}

Since $c$ is an e-class, it may include multiple equivalent \textit{Expr}, therefore we test whether any of the interpretable constraints are members of $c$. We define \textit{Constr} $\subseteq$ \textit{Expr}, denoting the generalized set of constraints appearing on the right of the if statements in (\ref{eqn:assume_abstraction_case}). An $\texttt{ASSUME}$ with a set of constraints represents a further restriction of the domain via additional intersections. To demonstrate, suppose $\abstraction{\texttt{x}} = [-3,3]$ then,
\[\abstraction{\texttt{ASSUME}(\texttt{x, x>0})} = [-3,3] \cap (0, \infty) = [1,3].\]
Examples in Section \ref{subsec:enable_equivs} demonstrate how this theory enables additional optimizations.
Our experience shows that, when combined with the rewriting described in Section \ref{sect:implementation}, it is only necessary to reason about the limited and computationally-efficient set of constraints described in (\ref{eqn:assume_abstraction_case}).

%%%%%%%%%%%%%%%%%%%%%%%%%%%%%%%%%%%%%%%%%%%%%%%%%%%%
% IMPLEMENTATION
%%%%%%%%%%%%%%%%%%%%%%%%%%%%%%%%%%%%%%%%%%%%%%%%%%%%
\section{Implementation} \label{sect:implementation}
We applied the theory described above to an RTL optimization tool that is built on the \texttt{egg} e-graph library~\cite{Willsey2021Egg:Saturation}. We optimize hardware designs at the RTL level of abstraction, operating on combinational logic on unsigned bitvectors. 
The tool parses input (System) Verilog using Yosys and sv2v~\cite{Wolf2013Yosys-ASuite}, converting it into an e-graph with bitwidth annotations, following the approach in~\cite{Coward2022AutomaticE-Graphs}. 
A set of parameterized and generalized constraint-aware rewrites at the word level is developed for this work. For concision of notation we exclude bitwidth annotations when describing rewrites in this paper. An online repository\footnote{https://figshare.com/s/e3ab2850662d24991cbc} summarizes rewrites described in this paper. 
Rewrites are automatically applied to the e-graph for a number of iterations, then a delay optimized expression is extracted from which a System Verilog implementation is generated. 

%%%%%%%%%%%%%%%%%%%%%%%%%%%%%%%%%%%%%%%%%%%%%%%%%%%%
% BITWIDTH REDUCTION
%%%%%%%%%%%%%%%%%%%%%%%%%%%%%%%%%%%%%%%%%%%%%%%%%%%%
\subsection{Bitwidth Reduction}
In RTL, expressions are evaluated over unsigned bit-vectors, therefore arithmetic is computed with respect to some modulo. When propagating finite unions of integer intervals, we use a conservative approximation to modular intervals.
\begin{equation}
    [l, u] \mod p = 
    \begin{cases}
       [l \mod p, u\mod p] &\text{if } \left\lfloor \frac{l}{p} \right\rfloor== \left\lfloor \frac{u}{p}\right\rfloor\\
       [0, p-1] &\text{else.}\\
    \end{cases}
\end{equation}

By propagating finite unions of integer intervals throughout the e-graph, corresponding to each classes' possible outputs, we enable bitwidth reduction. We maintain a bitwidth for each operand in the internal representation and are able to shrink this if we discover that the values which that operand can take would be representable in a smaller bitwidth. When combined with the $\ass$ node abstraction described above and the rewrites described in Section \ref{subsec:enable_equivs} below, we generate tighter approximations throughout, meaning that bitwidths can be squeezed to their minimum required precision.

%%%%%%%%%%%%%%%%%%%%%%%%%%%%%%%%%%%%%%%%%%%%%%%%%%%%
% ENABLING SUB-DOMAIN EQUIVALENCES IN E-GRAPHS
%%%%%%%%%%%%%%%%%%%%%%%%%%%%%%%%%%%%%%%%%%%%%%%%%%%%
\subsection{Enabling Sub-Domain Equivalences in E-Graphs}\label{subsec:enable_equivs}
We introduced the $\ass$ operator above and its abstraction. Focusing on RTL optimization, Table \ref{tab:assume_ax} describes rewrites to create, propagate and exploit $\ass$s. The $\ass$s will be induced by mux statements via the first rewrite in Table \ref{tab:assume_ax}. Recall that $\ass$s allow us to express multiple equivalence relations in the e-graph. We use rewrites of the form $\ass(x,c) \rightarrow \ass(y,c)$ to encode $x \cong_c y$. 

The need for $\ass$ nodes is illustrated via an example, \texttt{a==0 ? a : -a}, equivalent to, \texttt{a==0 ? 0 : -a}. Naively applying \texttt{a}$\rightarrow$\texttt{0}, to the e-graph merges the {\em non-equivalent} expression, \texttt{a==0 ? 0 : -0}, into the e-graph. Using Table \ref{tab:assume_ax} and the rewrite $\ass(\texttt{a, a==0}) \rightarrow \ass(\texttt{0, a==0})$, we produce the optimal expression.
\begin{align*}
&\texttt{(a==0)?}\texttt{a}\,: \texttt{-a}\rightarrow\\
&\texttt{(a==0)?}\ass(\texttt{a,a==0})\,: \ass(\texttt{-a,} \sim (\texttt{a==0}))\rightarrow\\
&\texttt{(a==0)?}\ass(\texttt{0,a==0})\,: \ass(\texttt{-a,} \sim (\texttt{a==0}))
\end{align*}
By construction we can treat $\ass$s as assignment statements in the implementation phase, meaning that they can be ignored in the optimized expression generated after rewriting. 

\begin{table}
    \centering
    \caption{$\ass$ node rewrites, describing creation, propagation and simplifications. The \texttt{op} operator can represent any operator defined in the intermediate language.}
    \begin{tabular}{l|l}
    Left-Hand Side & Right-Hand Side \\
    \hline
    $a\,?\,b\,:\,c$ & $a\,?\,\texttt{ASSUME}( b,\; a)\,:\,\texttt{ASSUME}(c,\; \sim a)$\\
    $\texttt{ASSUME}((a\; \texttt{op}\; b),\; c)$ & $\texttt{ASSUME}( a, \;c)\; \texttt{op}\;     \texttt{ASSUME}(b, \;c)$ \\
    $\texttt{ASSUME}(\texttt{ASSUME}(a,b),\; c)$ & $\texttt{ASSUME}(a,\; b \cup c)$ \\
    $\texttt{ASSUME}((a\,?\,b\,:\,c),\; a)$ & $\texttt{ASSUME}(b,\;a)$ \\
    $\texttt{ASSUME}((a\,?\,b\,:\,c),\; \sim a)$ & $\texttt{ASSUME}(c,\;\sim a)$
    \end{tabular}
    \label{tab:assume_ax}
\end{table}

The remainder of Table \ref{tab:assume_ax} is valuable in more complex scenarios. The second rewrite propagates $\ass$ nodes down through the e-graph towards the inputs, essential because the constrained sub-expression may occur at any point in the tree (or not at all). The third combines nested $\ass$ nodes by combining the two sets of conditions, which could be implemented as an AND of all the conditions. The final two rewrites prune unreachable branches. We discover the validity of additional sub-domain rewrites by evaluating the abstractions of the $\ass$ nodes and propagating this knowledge upwards through the e-graph.

In Section \ref{subsect:sub_domain equivs} we described a C expression. We now have a sequence of rewrites to prove the desired equivalence.
\begin{align*}
    \texttt{(x>0)?}&\texttt{fabs}(\texttt{x}) \hspace{6.6em}: \texttt{0}\hspace{8.48em}\to \\
    \texttt{(x>0)?}&\texttt{ASSUME}(\texttt{fabs}(\texttt{x}),\texttt{x>0}) 
                   : \texttt{ASSUME}(\texttt{0}, \sim(\texttt{x>0}))\to\\
    \texttt{(x>0)?}&\texttt{fabs}(\texttt{ASSUME}(\texttt{x},\texttt{x>0})) 
                   : \texttt{ASSUME}(\texttt{0}, \sim(\texttt{x>0}))\to\\
    \texttt{(x>0)?}&\texttt{ASSUME}(\texttt{x},\texttt{x>0}) 
                   \hspace{3.2em}: \texttt{ASSUME}(\texttt{0}, \sim(\texttt{x>0}))
\end{align*}
$\texttt{fabs}(\texttt{ASSUME}(\texttt{x},\texttt{x>0})) \to \texttt{ASSUME}(\texttt{x},\texttt{x>0})$ is proven valid via (\ref{eqn:assume_abstraction_case}), as $\abstraction{\texttt{ASSUME}(\texttt{x},\texttt{x>0})} = \abstraction{\texttt{x}} \cap (0, \infty)$. 

%%%%%%%%%%%%%%%%%%%%%%%%%%%%%%%%%%%%%%%%%%%%%%%%%%%%
% CONDITION REWRITING
%%%%%%%%%%%%%%%%%%%%%%%%%%%%%%%%%%%%%%%%%%%%%%%%%%%%
\subsection{Condition Rewriting} \label{subsect:cond_rewrite}
In Section \ref{subsect:relation_domains_egraphs} we described condition rewriting as a technique to mitigate the restrictions imposed by (\ref{eqn:assume_abstraction_case}). Using the rewrites described in Table~\ref{tab:condition_rewrites}, the tool attempts to transform $c\rightarrow c'$, where $c\in \textit{Expr}$ and $c' \in \textit{Constr}$.

\begin{table}
    \centering
    \caption{Condition rewrites, used to transform members of \textit{Expr} into members of \textit{Constr}.}
    \begin{tabular}{l| l}
    Transformation Rules & Inversion Rules \\
    \hline
    $a<b      \rightarrow a-b<0$     & $\sim (a=b)     \rightarrow a\neq b$  \\
    $a\leq b  \rightarrow a < b + 1$ & $\sim (a>b)     \rightarrow a \leq b$ \\  
    $a>b      \rightarrow a-b>0$     & $\sim (a\geq b) \rightarrow a < b$    \\ 
    $a\geq b  \rightarrow a>b -1 $   & $\sim (a< b)    \rightarrow a \geq b$ \\ 
    $a=b      \rightarrow a-b=0$     & $\sim (a\leq b) \rightarrow a > b$    \\ 
    $a=b      \rightarrow 0=b-a$     &          \\

    \end{tabular}
    \label{tab:condition_rewrites}
\end{table}

Consider an e-graph containing $\ass(\texttt{a-b, a>b})$. $\texttt{a>b}\notin \textit{Constr} \Rightarrow\abstraction{\ass(\texttt{a-b, a>b})} = \abstraction{\texttt{a-b}}$. By rewriting $\texttt{a>b} \rightarrow \texttt{a-b>0}$, we merge $\texttt{a-b>0}\in \textit{Constr}$ into the constraint e-class, triggering a refinement via (\ref{eqn:assume_abstraction_case}), $\abstraction{\ass(\texttt{a-b, a-b>0})} = \abstraction{\texttt{a-b}}\cap (0, \infty)$.

An e-class can contain many equivalent representations of a constraint so there is no need to find the single ideal representation. At the same time, the tool is also rewriting the expression under evaluation for optimization purposes. But a side benefit is that it may also discover how the particular imposed constraint impacts the expression.

In RTL design we often encounter conjunctions or disjunctions of conditions. We handle logical and/or via mux rewrites.
\begin{align}
    (a \land b) \, ? \, c \, : \, d &\rightarrow a\, ? \, (b \, ? \, c \,:\, d)\, : \, d \\
    (a \lor  b) \, ? \, c \, : \, d &\rightarrow a\, ? \, c\, : \, (b \, ? \, c \,:\, d)
\end{align}
These rewrites break conjunctions and disjunctions into simpler \textit{Expr} that the tool can reason about. Rewriting mux operations further mitigates the restrictions imposed by (\ref{eqn:assume_abstraction_case}). 

%%%%%%%%%%%%%%%%%%%%%%%%%%%%%%%%%%%%%%%%%%%%%%%%%%%%
% DELAY MODELING
%%%%%%%%%%%%%%%%%%%%%%%%%%%%%%%%%%%%%%%%%%%%%%%%%%%%
\subsection{Delay Modeling}
The final e-graph contains many functionally equivalent implementations of the input RTL. In this work, we target maximal performance and extract the design with the shortest critical path delay. If multiple designs achieve identical delay, we extract the smallest area circuit amongst them. 

We take a similar approach to previous work on multiplier design for FPGAs using e-graphs and construct a theoretical model of delay~\cite{Ustun2022IMpress:HLS}. For each operator we compute an estimate based on a fixed component architecture for the total number of two-input gates on the operator's critical path as a function of operator precision. At each operator the total delay to the output is the maximum delay across all its children plus its own delay. Using a theoretical model enables efficient design space exploration and avoids long logic synthesis runtimes. 

We extract the best design from the e-graph using $\texttt{egg}$'s standard extraction algorithm~\cite{Willsey2021Egg:Saturation} combined with a delay/area weighted sum objective function. For a performance prioritized optimization, common sub-expressions are not a significant factor, therefore we did not take an integer linear programming approach to extraction as elsewhere~\cite{Coward2022AutomaticE-Graphs}.

%%%%%%%%%%%%%%%%%%%%%%%%%%%%%%%%%%%%%%%%%%%%%%%%%%%%
% CASE STUDY - FP SUBTRACT
%%%%%%%%%%%%%%%%%%%%%%%%%%%%%%%%%%%%%%%%%%%%%%%%%%%%
\section{Case Study: Floating-Point Subtract} \label{sect:case_study}
To demonstrate the capabilities of such an approach we used the tool to automatically optimize a hardware implementation of a floating-point subtractor. These subtractors are amongst the most well studied hardware components and are the target of deep optimization efforts. Specifically we will demonstrate how the tool is able to optimize a half-precision floating-point subtractor, that computes $2^{ea}\times 1.ma - 2^{eb}\times 1.mb$, producing a half-precision floating-point output. We focus on the subtraction case because it is well-known to be harder~\cite{Beaumont-Smith1999ReducedArchitectures} due to the potential for cancellation. For simplicity, we consider the case where the output is rounded towards zero. We ignore exception handling and denormal, $\texttt{NaN}$ and infinite inputs.

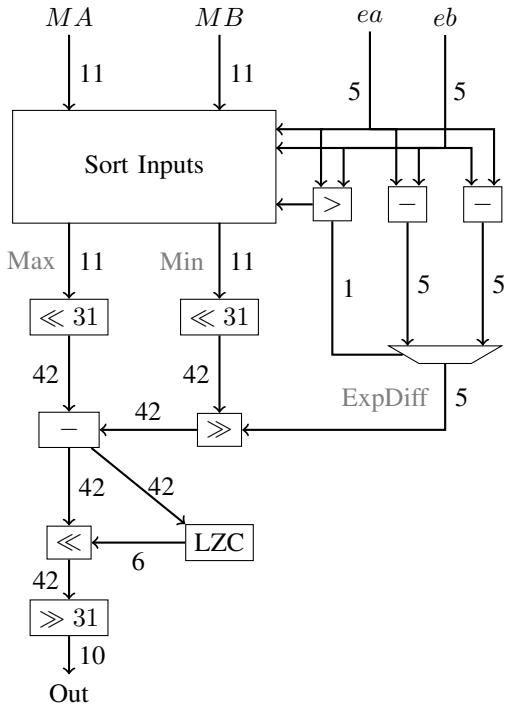
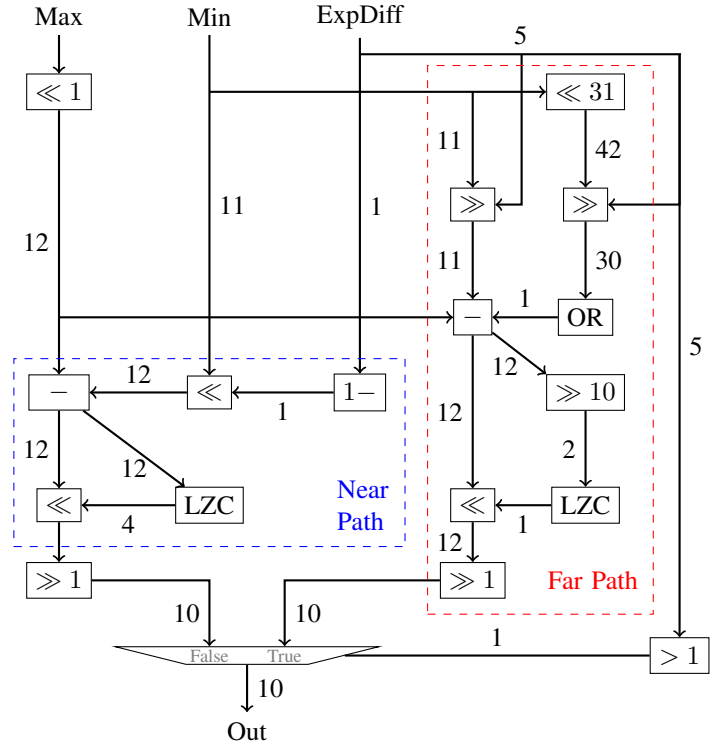
\begin{figure*}
\centering
\begin{subfigure}{.38\textwidth}
  \centering
  \begin{tikzpicture}

%draw input nodes
\node [shape=rectangle] at (1,21) (MA) {$MA$};
\node [shape=rectangle] at (3,21) (MB) {$MB$};
\node [shape=rectangle] at (5,21) (ea) {$ea$};
\node [shape=rectangle] at (6,21) (eb) {$eb$};
\node [] at (4.5,19.5) (ea_below) {};
\node [] at (6,19.25) (eb_below) {};

% exponent diff
\node [shape=rectangle,draw = black] at (4.5,18.5) (ea_gt_eb) {$>$};
\node [shape=rectangle,draw = black] at (5.5,18.5) (ea_sub_eb) {$-$};
\node [shape=rectangle,draw = black] at (6.5,18.5) (eb_sub_ea) {$-$};
\node [shape=trapezium,draw = black, minimum width = 1.5cm, trapezium stretches, shape border rotate = 180] at (6,16.5) (exp_diff) {};

% Exp diff arrows
\draw [-, thick] (ea) edge (5,19.5);
\node [] at (4.8, 20) {5};
\draw [->, thick] (5,19.5) -| (4.35,18.73);
\draw [->, thick] (5,19.5) -| (5.35,18.75);
\draw [->, thick] (5,19.5) -| (6.65,18.75);

\draw [-, thick] (eb) edge (6,19.25);
\node [] at (6.2, 20) {5};
\draw [->, thick] (6,19.25) -| (4.65,18.73);
\draw [->, thick] (6,19.25) -| (5.65,18.75);
\draw [->, thick] (6,19.25) -| (6.35,18.75);

\draw [-,thick] (ea_gt_eb) |- (exp_diff) node[pos=0.25, right] {1};
\draw [->,thick] (ea_sub_eb) -> (5.5,16.62) node[midway, right] {5};
\draw [->,thick] (eb_sub_ea) -> (6.5,16.62) node[midway, right] {5};

\draw [->, thick] (ea_gt_eb) edge (3.75, 18.5);
\draw [->, thick] (ea_below) edge (3.75, 19.5);
\draw [->, thick] (eb_below) edge (3.75, 19.25);

\node[gray] at (5.2, 15.9) {ExpDiff};
\node[gray] at (0.5, 17.75) {Max};
\node[gray] at (2.5, 17.75) {Min};

% Mantissa sorting
\node [shape=rectangle,draw = black, minimum width = 3.5cm, minimum height = 1.5cm] at (2,19) (sort_mant) {Sort Inputs};
\draw [->, thick] (MA) -> (1,19.75) node[midway, right] {11};
\draw [->, thick] (MB) -> (3,19.75) node[midway, right] {11};

%Alignment
\node [shape=rectangle,draw = black] at (1,17) (align_0) {$\ll 31$};
\node [shape=rectangle,draw = black] at (3,17) (align_1) {$\ll 31$};
\node [shape=rectangle,draw = black] at (3,15.5) (align_2) {$\gg$};

\draw[->,thick] (exp_diff) |- (align_2) node[pos=0.25,right] {5};
\node[] at (3,18.38) (mant_min) {};
\node[] at (1,18.38) (mant_max) {};
\draw[->,thick] (mant_min) -> (align_1) node[midway, right] {11};
\draw[->,thick] (mant_max) -> (align_0) node[midway, right] {11};
\draw[->,thick] (align_1) -> (align_2) node[pos=0.5,left] {42};

% Main Subtractor
\node [shape=rectangle,draw = black, minimum width = 0.8cm] at (1,15.5) (sub) {$-$};
\draw[->,thick] (align_2) -> (sub) node[pos=0.5,above] {42};
\draw[->,thick] (align_0) -> (sub) node[pos=0.5,left] {42};

% Normalization
\node [shape=rectangle,draw = black] at (3,14) (lzc) {LZC};
\node [shape=rectangle,draw = black] at (1,14) (renorm) {$\ll$};
\draw[->,thick] (sub) -> (lzc.north west) node[pos=0.5,right] {42};
\draw[->,thick] (lzc) -> (renorm) node[pos=0.5,below] {6};
\draw[->,thick] (sub) -> (renorm) node[pos=0.5,right] {42};

% Output
\node [shape=rectangle,draw = black] at (1,13) (right) {$\gg 31$};
\node [] at (1,12) (output) {Out};
\draw[->,thick] (renorm) -> (right) node[midway, left] {42};
\draw[->,thick] (right) -> (output) node[midway, right] {10};

\end{tikzpicture}
    \caption{Behavioural architecture, using a 42 bit subtraction.}
    \label{fig:naive_fp_add}
\end{subfigure}\quad%
\begin{subfigure}{.6\textwidth}
  \centering
  \begin{tikzpicture}

%draw input nodes
\node[] at (5,20) (exp_diff) {ExpDiff};
\node[] at (3,20) (mant_min) {Min};
\node[] at (1,20) (mant_max) {Max};
\node [shape=rectangle,draw = black] at (1,19) (align_0) {$\ll 1$};

%%%%%%%%%%%%%%%%%%%%%%%%%%%%%%%%%%%%%%%%%%%%%%%%%
% NEAR PATH
%%%%%%%%%%%%%%%%%%%%%%%%%%%%%%%%%%%%%%%%%%%%%%%%%
%Alignment
\node [shape=rectangle,draw = black] at (5,15) (sub_one) {$1-$};
\node [shape=rectangle,draw = black] at (3,15) (align_2) {$\ll$};

\draw[->,thick] (exp_diff) -> (sub_one) node[pos=0.5,right] {1};
% \draw[->,thick] (one) -> (sub_one);
\draw[->,thick] (sub_one) -> (align_2) node[pos=0.5,below] {1};

\draw[->,thick] (mant_max) edge (align_0);
\draw[->,thick] (mant_min) -> (align_2) node[pos=0.5,right] {11};

% Main Subtractor
\node [shape=rectangle,draw = black, minimum width = 0.8cm] at (1,15) (sub) {$-$};
\draw[->,thick] (align_2) -> (sub) node[pos=0.5,above] {12};
\draw[->,thick] (align_0) -> (sub) node[pos=0.5,left] {12};

% Normalization
\node [shape=rectangle,draw = black] at (3,13.5) (lzc) {LZC};
\node [shape=rectangle,draw = black] at (1,13.5) (renorm) {$\ll$};
\draw[->,thick] (sub) -> (lzc) node[pos=0.5,below] {12};
\draw[->,thick] (lzc) -> (renorm) node[pos=0.5,below] {4};
\draw[->,thick] (sub) -> (renorm) node[pos=0.5,left] {12};

% Output
\node [shape=rectangle,draw = black] at (1,12.5) (right) {$\gg 1$};

\draw[->,thick] (renorm) edge (right);

%%%%%%%%%%%%%%%%%%%%%%%%%%%%%%%%%%%%%%%%%%%%%%%%%
% FAR PATH
%%%%%%%%%%%%%%%%%%%%%%%%%%%%%%%%%%%%%%%%%%%%%%%%%
%Alignment
\node [shape=rectangle,draw = black] at (8,19) (shift_by_31) {$\ll 31$};
\node [shape=rectangle,draw = black] at (6.5,17.5) (shift_by_exp_0) {$\gg$};
\node [shape=rectangle,draw = black] at (8,17.5) (shift_by_exp_1) {$\gg$};
\node [shape=rectangle,draw = black] at (8,16) (or) {OR};

% Main Subtractor
\node [shape=rectangle,draw = black] at (6.5,16) (sub_far) {$-$};

% Normalization
\node [shape=rectangle,draw = black] at (8,15) (shfrt_10) {$\gg 10$};
\node [shape=rectangle,draw = black] at (8,13.5) (lzc_1) {LZC};
\node [shape=rectangle,draw = black] at (6.5,13.5) (normalize_far) {$\ll$};

\node [] at (4.88,19.5) (exp_diff_int_pt) {};
\node [] at (7.15,19.62) (exp_diff_int_pt_1) {};
\node [] at (9.25,19.62) (exp_diff_int_pt_2) {};
\node [] at (0.88,16) (max_mant_int_pt) {};
\node [] at (2.88,19) (min_mant_int_pt) {};
\node [] at (6.5,19.12) (min_mant_int_pt_1) {};

% Alignment and subtraction
\draw[->,thick] (min_mant_int_pt) edge (shift_by_31);
\draw[->,thick] (shift_by_31) -> (shift_by_exp_1) node[midway, right] {42};
\draw[->,thick] (shift_by_exp_1) -> (or) node[midway, right] {30};
\draw[->,thick] (shift_by_exp_0) -> (sub_far) node[midway, left] {11};
\draw[->,thick] (or) -> (sub_far) node[midway, above] {1};
\draw[->,thick] (min_mant_int_pt_1) -> (shift_by_exp_0) node[midway, left] {11};
\draw[-,thick] (exp_diff_int_pt) -> (9.27, 19.5) node[midway, above] {5};
\draw[->,thick] (exp_diff_int_pt_1) |- (shift_by_exp_0);
\draw[->,thick] (exp_diff_int_pt_2) |- (shift_by_exp_1);
\draw[->,thick] (max_mant_int_pt) -> (sub_far);

% Renorm
\draw[->,thick] (sub_far) -> (shfrt_10.north west) node[pos=0.8, left=0.1] {12};
\draw[->,thick] (shfrt_10) -> (lzc_1) node[midway, left] {2};
\draw[->,thick] (lzc_1) -> (normalize_far) node[midway, below] {1};
\draw[->,thick] (sub_far) -> (normalize_far) node[midway, left] {12};

% Output
\node [shape=rectangle,draw = black] at (6.5,12.5) (right_1) {$\gg 1$};
\draw[->,thick] (normalize_far) -> (right_1) node[midway,left] {12};

% Condition
\node [shape=rectangle,draw = black] at (9.25,11.5) (condition) {$> 1$};

% Output
\node [shape=trapezium,draw = black, minimum width = 3.5cm, trapezium stretches, shape border rotate = 180] at (3.5,11.5) (final_mux) {};
\node [] at (3.5,10.5) (output) {Out};
\draw[->,thick] (right_1) -| (4,11.625) node[pos=0.75, right] {10};
\draw[->,thick] (right) -| (3,11.625) node[pos=0.75, left] {10};

\draw[->,thick] (final_mux) -> (output) node[midway, right] {10};
\draw[thick] (condition) -> (4.8,11.5) node[midway, above] {1};
\draw[->, thick] (exp_diff_int_pt_2) -> (condition) node[midway, right] {5};

% Outline of Near & Far Path
\node[shape = rectangle, draw=blue, dashed, minimum width = 5.2cm, minimum height = 2.5cm] at (3, 14.2) {};
\node [text width = 1cm, color=blue] at (5.2, 13.5) {Near Path};

\node[shape = rectangle, draw=red, dashed, minimum width = 3cm, minimum height = 7.3cm] at (7.4, 15.7) {};
\node [text width = 2cm, color=red] at (8.5, 12.5) {Far Path};

\node[gray] at (3, 11.5) {\scriptsize False};
\node[gray] at (4, 11.5) {\scriptsize True};
\end{tikzpicture}
    \caption{Dual path optimized architecture, using two 12 bit subtractors. Near path uses a one bit alignment shift, followed by a larger renormalization stage. Far path uses a 12 bit alignment shift and a single bit renormalization.}
    \label{fig:opt_fp_add}
\end{subfigure}
\caption{Half-precision floating-point subtractor architectures. Input mantissas have the implicit one appended. Edge labels represent bitwidths. The inputs are sorted according to $ea>eb \lor (ea == eb \land MA>MB)$. In the optimized architecture diagram, we omitted input sorting and exponent difference calculation blocks, as they were unchanged by the optimization.}
\label{fig:test}
\end{figure*}

The input design in Figure \ref{fig:naive_fp_add}, is a naive, easy to write, implementation of the mantissa calculation for floating-point subtraction. In the naive design, the mantissas are concatenated with preceding zeros to ensure that after the alignment, all bits are retained. This results in a 42 bit subtractor, the result of which is normalized, using an LZC and a right shift. Discarding the leading one from this result, the output mantissa is then the 10 most significant remaining bits.

The most well-known floating-point subtract optimization known as the near-path/far-path optimization, stems from the observation that the critical path is never fully exercised~\cite{Beaumont-Smith1999ReducedArchitectures,Farmwald1981OnUnits}. It splits the design into two paths. The near path is taken when $|ea-eb|< 2$, requiring only a small alignment shift. The far path is taken when $|ea-eb|>1$. On this path catastrophic cancellation cannot occur, simplifying the renormalization logic. Figure \ref{fig:lzc_diagram} shows a related optimization. 
%\gc{Is this the right place for the following sentence?} To automate this optimization we require accurate bounds on the intermediate signal values and the ability to exploit sub-domain equivalences. 

The tool parses the RTL corresponding to Figure \ref{fig:naive_fp_add} and generates an initial e-graph. Typically LZCs are expressed in Verilog as case statements. To enable operator specific rewrites, we include an LZC operator in the intermediate language, with accompanying rewrites to map appropriate case statements to this operator. 
%\gcc{We pre-process the e-graph with one rewriting iteration to convert suitable case statements to LZC operators, then prune the case statement to simplify the e-graph, which avoids unnecessary exploration.}{(suggest deletion -- too detailed and short of space?)}

Once the tool has mapped the input design to its intermediate language optimization begins. Firstly, the tool introduces the case split into the e-graph via a rewrite, which is intended to possibly isolate catastrophic cancellation to a single branch.
\[a - (b>>c) \rightarrow (c>1) \, ? \, a - (b>>c)\, :\, a - (b>>c)\]
Note that this inserts a mux directly after the subtraction, which may be beneficial in some instances, but in this case using mux propagation rewrites, the tool pushes the mux towards the output.
\begin{equation*}
    a \texttt{ op } (b \, ? \, c \, : \, d) \rightarrow b \, ? \, a \texttt{ op } c \, : \, a \texttt{ op } d
\end{equation*}
This duplicates operations, leaving a mux at the output between two identical branches.
%\begin{align}
%    \lsub{42}{\mathrm{Max0}} &= {\{\lsub{11}{\mathrm{Max}},\lsub{31}0\}},\; %\lsub{42}{\mathrm{Min0}} = {\{\lsub{11}{\mathrm{Min}},\lsub{31}0\}}\\
%    \lsub{42}{\mathrm{Sum}} &= \lsub{42}{\mathrm{Max0}} - \lsub{42}{\left(\lsub{42}{\mathrm{Min0}} %\gg \lsub{5}{\mathrm{ExpDiff}}\right)}\\
%    \lsub{10}{\mathrm{Branch}} & = \lsub{42}{\left(\lsub{42}{\mathrm{Sum}} \ll %\lsub{6}{\mathrm{LZC}(\lsub{42}{\mathrm{Sum}})}\right)} \gg 31\\
%    \lsub{10}{\mathrm{Out}}&=  \lsub{5}{\mathrm{ExpDiff}} > 1 \, ? \,\lsub{10}{\mathrm{Branch}}\,
%    : \lsub{10}{\mathrm{Branch}}.
%\end{align}
Using the rewrites from Table~\ref{tab:assume_ax}, the tool creates and propagates $\ass$s down each branch and after several rewriting iterations, the following are created.
\begin{align}
\ass&(\mathrm{ExpDiff}, \hspace{1.4em}\mathrm{ExpDiff} > 1) \label{eqn:first_exp_assume}  \\
\ass&(\mathrm{ExpDiff}, \sim(\mathrm{ExpDiff} > 1)) \label{eqn:second_exp_assume}
\end{align}
Computing the abstraction of (\ref{eqn:first_exp_assume}), (\ref{eqn:assume_abstraction_case}) takes effect immediately. For (\ref{eqn:second_exp_assume}), two sequential condition rewrites transform it into an equivalent \textit{Constr}, $\mathrm{ExpDiff} < 2$.
These constrained value ranges are propagated along each branch triggering a chain of branch specific rewrites and bitwidth reductions. 

After 11 iterations of rewriting an e-graph of approximately 40,000 nodes and 14,000 classes is grown.
The optimized design shown in Figure \ref{fig:opt_fp_add} is extracted from this e-graph, within a total runtime of 22 minutes, the majority of which is spent growing the e-graph. 
Note that the two branches are reduced to distinct implementations as a result of the constraint-aware rewrites.
Further optimizations apply to even this design, but these are often gate level optimizations, a level of optimization we have left to the logic synthesis tools. 

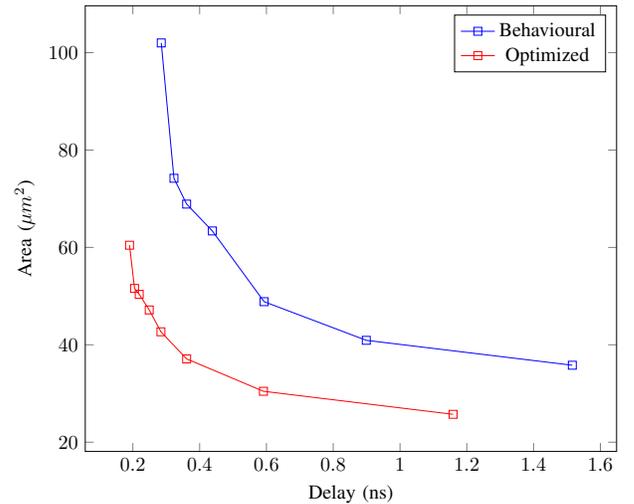
\begin{figure}
    \centering
    \begin{tikzpicture}[scale=0.75]
\begin{axis}[
    xlabel ={Delay (ns)},
    ylabel ={Area ($\mu m^2$)}
    ]
\addplot[
    color=blue,
    mark=square,
    ]
    coordinates {
        (0.285, 101.99808)
        (0.323, 74.20032 )
        (0.361, 68.89248 )
        (0.438, 63.37944 )
        (0.593, 48.85128 )
        (0.899, 40.91688 )
        (1.516, 35.80056 )
        % (2.748, 31.86072 )
    };
    \addlegendentry{Behavioural}
    \addplot[
    color=red,
    mark=square,
    ]
    coordinates {
        (0.19 , 60.45192)
        (0.205, 51.5736 )
        (0.219, 50.35608)
        (0.249, 47.15496)
        (0.284, 42.66792)
        (0.361, 37.10016)
        (0.591, 30.45168)
        (1.159, 25.7184 )
    };
    \addlegendentry{Optimized}
\end{axis}
\end{tikzpicture}
    \caption{Area-delay plot of competing floating-point subtractors.}
    \label{fig:fp_sub_area_delay}
\end{figure}

The input and optimized RTLs are proven equivalent using the Synopsys Datapath Validation (DPV) tool, a formal equivalence checking tool that runs in minutes on this problem. We synthesized both designs at a range of delay targets using Synopsys Fusion Compiler for a TSMC 7nm cell library. The results are shown in Figure \ref{fig:fp_sub_area_delay}. The optimized design can achieve a 33\% lower delay in a circuit area 41\% smaller than the behavioural architecture.

%%%%%%%%%%%%%%%%%%%%%%%%%%%%%%%%%%%%%%%%%%%%%%%%%%%%
% RESULTS
%%%%%%%%%%%%%%%%%%%%%%%%%%%%%%%%%%%%%%%%%%%%%%%%%%%%
\section{Further Results} \label{sect:results}
To demonstrate that the results of Section \ref{sect:case_study} are not hand-tuned to the case study, this section uses the tool to optimize several different design. 
We ran the tool for six iterations on smaller test cases generating e-graphs of less than 150 nodes, running in under 0.25 seconds. We demonstrate how the tool can automatically do dead code elimination and generalize optimizations learnt from the floating-point case study. The behavioural and optimized RTLs are proven equivalent using the DPV tool. The competing RTLs are synthesized using Fusion Compiler at the minimum delay target that each implementation can meet. Table \ref{tab:synth_results} summarizes the results. 

\begin{table}
    \centering
    \caption{Logic synthesis results using Fusion compiler for circuit delay (ns) and circuit area ($\mu m^2$). Each implementation is synthesized at the minimum delay target it can achieve. }
    \begin{tabular}{|c|c|r|l|r|}
         Test Case       & \multicolumn{2}{c|}{Behavioural} &  \multicolumn{2}{c|}{Optimized} \\
                        & ns & $\mu m^2$   & \multicolumn{1}{c|}{ns} & \multicolumn{1}{c|}{$\mu m^2$} \\
        \hline
        FP Sub         & 0.285      &  102.0           & 0.190 (\textbf{-33\%})    & 60.4 (\textbf{-41\%})\\
        float to unorm   & 0.055      & 17.6             & 0.056 (\textbf{+2\%})     & 13.6 (\textbf{-23\%})\\
        interpolation  & 0.245      &  433.0           & 0.254 (\textbf{+3\%})     & 353.0 (\textbf{-18\%})\\
        unorm to float & 0.039      &  13.4            &  0.039 (\textbf{+0\%})   & 7.0 (\textbf{-48\%}) \\
    \end{tabular}
    \label{tab:synth_results}
\end{table}

The float to unorm design converts a half-precision float (less than or equal to 1 in magnitude)  to a unorm11, rounding down, as described in the DirectX specifications~\cite{Microsoft2022DirectX-Specs}. The tool reuses the round-off based optimizations from Section \ref{sect:case_study}. The interpolation example is a kernel from an Intel media module, computing an interpolation between four pixels and clamping the output. For certain clamping thresholds, the tool automatically detects that the threshold can never be met and optimizes the clamping away. 
\[c \;? \; a\; :\; b \rightarrow b \;\textrm{ if } \abstraction{c} == [0,0]\]
This test case relies on rewriting to obtain tight approximations to the range of outputs and only with this rewriting can the tool prove that the clamping is unnecessary. Namely, naive interval arithmetic would not suffice.

The unorm to float design special cases zero inputs, such that they are handled on a separate path. The tool automatically propagates the domain restriction and applies the constraint-aware optimizations generating a smaller circuit that matches the behavioural's performance. 
In this example the interplay of rewriting and program analysis is invaluable.

%%%%%%%%%%%%%%%%%%%%%%%%%%%%%%%%%%%%%%%%%%%%%%%%%%%%
% CONCLUSION
%%%%%%%%%%%%%%%%%%%%%%%%%%%%%%%%%%%%%%%%%%%%%%%%%%%%
\section{Conclusion}
This paper combines e-graphs with constraint-aware program analysis to automate RTL optimization exceeding the capabilities of existing EDA tools. 
By representing multiple equivalence relations in an e-graph we exploit branch specific optimizations and compute tight approximations to intermediate signals using an extension of interval arithmetic. These techniques enable bitwidth reduction, dead code elimination and automated case splitting. We automated the optimization of a floating-point subtraction unit recreating efficient human implementations and saving 41\% of area and 33\% of delay. In other test cases the tool reduced circuit area by up to 48\% with minimal delay penalty demonstrating its generalizability.

Future work will look to refine the delay model by considering a bit-level delay model, and explore multi-objective optimization to generate Pareto curves of designs. Having matched human designers on floating-point unit design, we hope to use the tool to discover novel architectures in the near future. An unexplored interactive tool usage, would be for designers to propose case-splits based on their intuition and have the tool automatically optimize the proposed design. 

\section*{Acknowledgment}
The authors acknowledge the discussions with the creators of the \texttt{egg} library, in particular Pavel Panchekha and Max Willsey for the initial idea to use a variation on our \texttt{ASSUME} nodes.

\bibliographystyle{IEEEtran}
\bibliography{references.bib}

\end{document}